\documentclass[aps,twocolumn,superscriptaddress,nofootinbib]{revtex4-1}
\usepackage{graphicx}
\usepackage{hyperref}
\usepackage{amsmath,amssymb}
\pdfoutput=1

\begin{document}

\newcommand{\FIRSTAFF}{\affiliation{Department of Physics,
			University at Buffalo, SUNY
			Buffalo,
			NY 14260
			USA}}
	
\author{Wei-Chen Lin}
\email{weichenl@buffalo.edu} 
\author{William H. Kinney}
\email{whkinney@buffalo.edu} 
\FIRSTAFF
	
\date{\today}

\title{The Trans-Planckian problem in Tachyacoustic Cosmology }

\begin{abstract}
We study Tachyacoustic models of cosmology, for which a scale-invariant perturbation spectrum is generated via superluminal sound speed instead of accelerated expansion, as in the case of inflation. We derive two bounds on the size of acoustic horizon which constrain the duration of tachyacoustic evolution, and therefore generation of primordial perturbations. We show that existing models cannot solve the horizon problem without violating the condition that all physical scales -- such as the Hubble parameter, the pressure, and the length scale at which quantum modes freeze out and become classical -- be sub-Planckian. 
\end{abstract}


\maketitle

\section{Introduction \label{sec1}}
Inflationary cosmology \cite{Starobinsky:1980te,Sato:1981ds,Sato:1980yn,Kazanas:1980tx,Guth:1980zm,Linde:1981mu,Albrecht:1982wi} is the most successful and widely accepted theory of the very early universe. From the observational point of view, however, inflationary cosmology is not the unique theory able to explain the origin of the primordial density fluctuations, which seed the Cosmic Microwave Background (CMB) anisotropy and structure formation in the universe.   
Refs. \cite{Geshnizjani:2011dk,Geshnizjani:2011rm,Geshnizjani:2013lza,Geshnizjani:2014bya} derived general conditions on cosmology for the generation of primordial perturbations consistent with observation, and found four general classes of models, assuming standard General Relativity:
\begin{enumerate}
	\item{A period of accelerated expansion ({\it i.e.} inflation).}
	\item{A speed of sound faster than the speed of light.}
	\item{Violation of the Null Energy Condition.}
	\item{Inherently quantum-gravitational physics.}
\end{enumerate} 

In this paper, we review a model of the second kind, a type of k-essence model dubbed  \textit{tachyacoustic cosmology} proposed by Bessada et al. \cite{Bessada:2009ns} based on the work by Magueijo \cite{Magueijo:2008pm}.  As an alternative to inflation, tachyacoustic cosmology was proposed to solve the horizon problem and creates a nearly scale invariant power spectrum via a superluminal speed of sound, $ c_S >1 $. (Although this special type of model involves superluminal sound speed, it has been argued that such fields do not violate causality and are as consistent as models with subluminal speed of sound \cite{Babichev:2007dw}.) In this paper, we consider dynamics in tachyacoustic models which are consistent with present observational constraints. We find that for such models to satisfy the horizon problem, we unavoidably need to introduce a period when the field also violates the condition that all scales be sub-Planckian. Therefore, existing tachyacoustic models are not self-consistent.  Our analysis differs from that of Refs. \cite{Geshnizjani:2011dk,Geshnizjani:2011rm,Geshnizjani:2013lza,Geshnizjani:2014bya}, because we consider not only the requirement of scale invariance, but also the \textit{normalization} of the primordial power spectrum. Our analysis is, however less general, since we specialize to the power-law models proposed in Ref. \cite{Bessada:2009ns}.

The rest of this paper is organized as follows. In Sec. II, we briefly review tachyacoustic cosmology. In Sec. III, we show that the horizon problem cannot be solved in a general tachyacoustic cosmology setting without violating the sub-Planckian condition. In Sec. IV, we show that the thermal Tachyacoustic Cosmology is also facing the same issue. The conclusion is in Sec. V.

\section{Tachyacoustic Cosmology \label{sec2}}

We begin this section with a brief review of tachyacoustic cosmology \cite{Bessada:2009ns}, based on the generalization of the inflationary flow formalism \cite{Kinney:2002qn} introduced by Bean et al. \cite{Bean:2008ga}. In Ref. \cite{Bessada:2009ns}, the authors utilize the generalized flow formalism with additional conditions on the background evolution to reconstruct the corresponding Lagrangians and study exact tachyacoustic solutions. In the following, we only review the background evolution of tachyacoustic cosmology since it is the part relevant to the issue addressed in this paper. One can find more details of the reconstruction in Refs. \cite{Bean:2008ga, Bessada:2009ns}, and especially about the issue of field redefinition in Ref. \cite{Lin:2018edm}.  

As a type of k-essence models, the Lagrangian of a tachyacoustic field has the general form $ \mathcal{L}= \mathcal{L}[X,\phi]$, where $ X \equiv \frac{1}{2} g^{\mu \nu} \partial_{\mu}\phi  \partial_{\nu}\phi$ is the canonical kinetic term. Homogeneous modes of this scalar field form a perfect fluid with energy-momentum tensor
\begin{equation}
\label{PF_energy_momentum_tensor}
T_{\mu \nu}=(p+\rho)u_{\mu}u_{\nu}-pg_{\mu\nu},
\end{equation}
with
\begin{align}
\label{PF_pressure}
p(X,\phi)&= \mathcal{L}(X,\phi),
\\
\label{PF_energy_density}
\rho(X,\phi)&= 2X\mathcal{L}_X-\mathcal{L}(X,\phi),
\\
\label{fluid_velocity}
u_{\mu}&= \frac{\partial_{\mu}\phi}{\sqrt{2X}},
\end{align}
where $ \mathcal{L}_{X} \equiv \partial \mathcal{L}/ \partial X $.
The speed of sound is non-trivially given by 
\begin{equation}
\label{sound_speed}
c^{2}_{S} \equiv \frac{p_X}{\rho_X}=\Bigg(1+2X\frac{\mathcal{L}_{XX}}{\mathcal{L}_X}\Bigg)^{-1},
\end{equation}
where the canonical limit $ \mathcal{L}_{XX}=0 $ leads to $ c_{S}=1 $. Then in this type of k-essence model, curvature perturbations freeze out and become classical at the \textit{acoustic} horizon, $c_S k = a H$, which corresponds to the Hubble length when $c_S = 1$, but otherwise is dynamically independent \cite{Garriga:1999vw}. If we allow for superluminal sound speed $c_S > 1$, it is possible generate super-Hubble perturbations without inflation. 

We will focus on the ``tachyacoustic'' cosmology models constructed in Ref. \cite{Bessada:2009ns}, which can be specified by a set of three parameters $ \{ \epsilon, s, \tilde{s} \} $ related to the background evolution of Hubble parameter $ H $, speed of sound $ c_S $, and the derivative of Lagrangian with respect to the canonical kinetic term, $ \mathcal{L}_{X} \equiv \partial \mathcal{L}/ \partial X $, as functions of the number of e-folds $ N $ during the tachyacoustic field dominated phase as
\begin{equation}
\begin{aligned}
\label{Tachy_Forms}
H & = H_{i} e^{-\epsilon N},
\\
c_S & = c_{S,i}e^{s N},
\\
\mathcal{L}_{X} &=  Ae^{-\tilde{s} N},
\end{aligned}
\end{equation}
where the convention $ N=\ln a/a_{i} $ with $ a_{i} $ as the scale factor at the beginning of the tachyacoustic phase is used. The parameter $ \epsilon $ is the first slow-roll parameter 
\begin{equation}
\label{epsilon}
\epsilon=\frac{-1}{H}\frac{dH}{dN}, 
\end{equation}
and the parameter $ s $ is the first flow parameter of speed of sound \cite{Peiris:2007gz, Kinney:2007ag}
\begin{equation}
\label{s_definition}
s=\frac{1}{c_S}\frac{dc_S}{dN}, 
\end{equation}
which we take to be constant for the solutions in Eq. (\ref{Tachy_Forms}). The linear scalar perturbations of the Tachyacoustic Cosmology follow the same result in a general k-essence model, in which a general quadratic action for the curvature perturbation $\zeta$ can be written in terms of the dynamical variable $d y \equiv c_S d\tau$ as \cite{Khoury:2008wj}
\begin{equation}
\label{general_quadratic_action}
S_2 = \frac{M_P^2}{2} \int{dx^3 dy q^2 \left[\left(\frac{d \zeta}{d y}\right)^2 - \left(\nabla \zeta\right)^2\right]},
\end{equation}
where $\tau$ is the conformal time, $ds^2 = a^2\left(\tau\right) \left[d \tau^2 - d {\bf x}^2\right]$, and $q$ is given by
\begin{equation}
\label{q_factor}
q \equiv \frac{a \sqrt{2 \epsilon}}{\sqrt{c_S}}. 
\end{equation}
By defining a canonically normalized scalar mode function $v \equiv M_{\rm P} q \zeta$, the associated mode equation becomes
\begin{equation}
\label{general_mode_equation}
v_k'' + \left(k^2 - \frac{q''}{q}\right) v_k = 0,
\end{equation}
where a prime denotes differentiation with respect to $y$. From Eqs. (\ref{Tachy_Forms}) and (\ref{q_factor}), one can obtain that when $ s=-2\epsilon $ and $ q \propto y^{-1} $,  Eq. (\ref{general_mode_equation}) reduces to 
\begin{equation}
\label{scale_invariant_mode_equation}
v_k'' + \left(k^2 - \frac{2}{y^{2}}\right) v_k = 0,
\end{equation}   
which admits the solution 
\begin{equation}
\label{scale_invariant_solution}
v_k=\frac{e^{-iky}}{\sqrt{2k}}(1-\frac{i}{ky}),
\end{equation}
 consistent with usual Bunch-Davies boundary condition,
\begin{equation}
\label{initial_condition_Kessence}
\lim_{y\to - \infty} v_k=\frac{e^{-iky}}{\sqrt{2k}}. 
\end{equation} 
This special solution corresponds to the de Sitter limit in a canonical inflationary model by the substitution $ \tau \rightarrow y $. Therefore, the power spectrum of the curvature perturbation is scale-invariant and given by 
\begin{equation}
\label{Scale_invariant_Power_Spec_Tachy}
\mathcal{P}_{\zeta}(k)= \lim_{ky\to 0^{-}}  \frac{k^3}{2\pi^2}\frac{|v_k|^2}{M_P^2 q^2}=\frac{(1+\epsilon)^2}{\epsilon}\frac{1}{8\pi^2}\frac{H^2}{M_P^2}\frac{1}{c_S},
\end{equation}
which is similar to the result in Ref. \cite{Magueijo:2008pm}. In fact, with constant flow parameters (\ref{Tachy_Forms}), the mode equation (\ref{general_mode_equation}) can be solved exactly and the power spectrum of the curvature perturbation is given by \cite{Bessada:2009ns}
\begin{equation}
\label{Power_Spec_Tachy}
\mathcal{P}_{\zeta}=\frac{|f(\nu)|^{2}}{8\pi^2}\frac{H^2}{M_P^2}\frac{1}{c_S \epsilon} \bigg\vert_{c_S k = a H},
\end{equation}
where the power spectrum amplitude is evaluated when the quantum mode crosses the acoustic horizon, $c_S k = a H$. Here $ f(\nu) $ and $ \nu $ are given by\footnote{We notice that in Ref. \cite{Bessada:2009ns}, $ \nu $ is mistakenly depends to the gauge parameter $ \tilde{s} $. The form given in Eq. (\ref{nu}) matches the result found in Ref. \cite{Bean:2008ga}. } 
\begin{equation}
\label{f(nu)}
f(\nu)=2^{\nu-\frac{3}{2}}\frac{\Gamma(\nu)}{\Gamma(3/2)}(1-\epsilon-s)^{\nu-\frac{1}{2}},
\end{equation}
and
\begin{equation}
\label{nu}
\nu=\frac{3-2s-\epsilon}{2(1-\epsilon-s)}.
\end{equation}
The scalar spectral index of perturbations for a tachyacoustic solution is given by 
\begin{equation}
\label{Tachya_tilt}
n_s=1-\frac{2\epsilon+s}{1-\epsilon-s}, 
\end{equation}
which has a scale-invariant limit, $ s=-2\epsilon $, and has been shown to be a dynamical attractor \cite{Bessada:2012zx}. In the scale-invariant limit, 
\begin{equation}
\label{Scale_invariant_f(nu)}
f(\nu)=1+\epsilon, 
\end{equation}
with $ \nu=3/2 $, Eq. (\ref{Power_Spec_Tachy}) reduces to the form of Eq. (\ref{Scale_invariant_Power_Spec_Tachy}). Notice that the results are independent of the gauge parameter $\tilde{s}  $, \textit{i.e.} in the linear perturbation level, all models in Tachyacoustic Cosmology are observationally indistinguishable.

Lastly, the condition of solving the horizon problem in  Tachyacoustic Cosmology is similar to that in k-inflation. If we consider a comoving wave mode with wavelength of order the horizon size today, $k_0 = (a_0 H_0)$, the condition of solving the horizon problem requires 
\begin{equation}
\label{general_horizon_problem}
\frac{c_{S}(a_i)}{a_i H_i} \geq (a_0 H_0)^{-1}, 
\end{equation}
where $ a_i $ is the scale factor at the beginning of the tachyacoustic phase.

\section{Super-Planckian density issue in Tachyacoustic Cosmology \label{sec3} }

\begin{figure}[h!]
	\centering
	\includegraphics[scale=0.8]{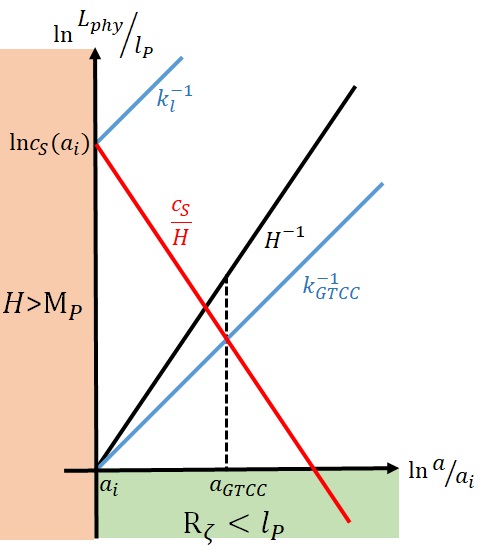}
	\caption{ A schematic picture of a scale-invariant tachyacoustic cosmology model saturating the sub-Planckian bound $ H(a_i) < M_P $. In this $ \ln \frac{L_{phy}}{l_P} $ vs $\ln \frac{a}{a_i}$ diagram, the evolution of the physical length of modes is given by the family of straight lines with slope equal to 1. The quantum-to-classical transition of fluctuations happens during the crossing of acoustic horizon $R_{\zeta}=\frac{c_S}{H}$ (red). $ H_{l} $ is the Hubble parameter when the longest mode $ k_l $, which exited the acoustic horizon at the beginning of tachyacoustic phase, enters the Hubble horizon. Solving the horizon problem requires $ H_{l} \leq H_0 $. The orange area is forbidden since it represents the part with super-Planckian energy density $ H>M_P $. Meanwhile, the green area represents the part that the acoustic horizon is shorter than the Planck length, so the quantum-to-classical transition of the perturbation is invalid. }  
	\label{fig:TC_GTCC}
\end{figure}

In this section we first show that in the radiation-dominated tachyacoustic model considered in Ref. \cite{Bessada:2009ns}, solving the horizon problem guarantees a period of super-Planckian energy density. We next show that solving the horizon problem in a more general tachyacoustic model, $ \epsilon \gg 1 $, still results the super-Planckian physical quantities. To simplify our analysis, we will approximate the late universe as having decelerating expansion throughout, ignoring late-time acceleration. This does not substantially effect our bounds, since we find that the longest wavelength modes consistent with sub-Planckian energies re-enter the horizon long before matter/radiation equality. In this case, the condition of solving the horizon problem (\ref{general_horizon_problem}) can be written as\footnote{More specifically, Eq. (\ref{H_l<H_0}) and Eq. (\ref{general_horizon_problem}) are equivalent if we ignore the fact that modes are exiting instead of entering the Hubble horizon due to the current accelerated expansion.}    
\begin{equation}
\label{H_l<H_0}
H_{l} \leq H_0,
\end{equation}  
where $ H_{l}  $ is the Hubble parameter when the mode having the size of the acoustic horizon at the onset of tachyacoustic phase enters the Hubble horizon (the mode $ k_l $ in Fig. \ref{fig:TC_GTCC}). Secondly, since the tilt of the scalar spectral index (\ref{Tachya_tilt})  is small from observational constraint \cite{Aghanim:2018eyx}, from Eq. (\ref{f(nu)}) one can easily show that the correction to Eq. (\ref{Scale_invariant_f(nu)}) is negligible. Therefore we use the scale invariant limit to simplify our analysis. 

Under the condition of scale invariance, we can relate the scale invariant power spectrum (\ref{Scale_invariant_Power_Spec_Tachy}) to the CMB measurement $ \mathcal{P}_{\zeta}(k)=\mathcal{P}_{\zeta} (k_{*}=0.05Mpc^{-1}) \sim 2 \times 10^{-9} $ \cite{Aghanim:2018eyx} as   
\begin{equation}
\label{Power_Spec_Tachy_CMB_normal_SI}
\mathcal{P}_{\zeta}=\frac{1}{8\pi^2}\frac{H^2}{M_P^2}\frac{(1+\epsilon)^2}{c_S \epsilon} \sim 2 \times 10^{-9},
\end{equation}
which gives
\begin{equation}
\label{TachyCos_H_constraint}
\left(\frac{H}{M_P}\right)^2 \sim 10^{-7} \frac{c_S\epsilon}{(1+\epsilon)^2}. 
\end{equation} 
To avoid quantum-gravitational effects we must have   
\begin{equation}
\label{H_i_limit}
\frac{H(a_i)}{M_P} < 1, 
\end{equation}
where $ a_i $ is the scale factor at the beginning of the Tachyacoustic Cosmology phase.  Eqs. (\ref{TachyCos_H_constraint}) and (\ref{H_i_limit}) then provide a maximal value of speed of sound the model can have without entering the super-Planckian region
\begin{equation}
\label{c_S_i_limit}
c_S(a_{i}) \lesssim \frac{10^{7}(1+\epsilon)^2}{\epsilon}, 
\end{equation}
where the upper bound is from the limit $ H(a_i) \sim M_P $. The crucial difference between the Tachyacoustic Cosmology and an inflationary scenario is that in inflation we can take $ \epsilon \ll 1 $ but in a Tachyacoustic Cosmology model we have  $ \epsilon >1 $.

We first consider the radiation-dominated tachyacoustic model proposed in Ref. \cite{Bessada:2009ns}, which is also the simplest case. Because of the radiation-dominated expansion both before and after the decay of the tachyacoustic field into standard model particles, $ H_l $ is independent of the decay time. By saturating the energy bound (\ref{H_i_limit}) and using Eq. (\ref{Tachy_Forms}) for $ c_S(N) $ and $ H(N) $ with $ s=-2\epsilon=-4 $, the relation between $ H_l $ and $ c_S(a_i) $ is given by 
\begin{equation}
\label{H_l_RD_eq}
\ln \frac{M_P}{H_l} =2 \ln c_S(a_i).  
\end{equation}
Meanwhile, with $ \epsilon=2 $, the bound on the speed of sound (\ref{c_S_i_limit}) is 
\begin{equation}
\label{c_S_i_limit_RD}
c_S(a_{i}) \lesssim 4.5 \times 10^{7}.  
\end{equation}
Then Eqs. (\ref{H_l_RD_eq}) and (\ref{c_S_i_limit_RD}) give a bound on $ H_l $ as
\begin{equation}
\label{H_l_value_RD}
H_l \gtrsim 5 \times 10^{-15}M_P, 
\end{equation}
which is significantly larger than the current Hubble parameter $ H_0 \approx 10^{-62}M_P $. Therefore, the horizon problem is unsolved by this model given the bound (\ref{H_i_limit}) since Eq. (\ref{H_l<H_0}) is violated --- that is, solving the horizon problem guarantees a period of super-Planckian energy density. In the next part, we show that even in a general tachyacoustic model with $ \epsilon \gg 1 $, solving the horizon problem unavoidably introduces the super-Planckian issue.   

Next we consider a more general tachyacoustic model, with arbitrary equation of state. Since the equation of state of a general tachyacoustic field is different from that of radiation, $ H_l $ depends on the energy scale of the decay of the tachyacoustic field, $ H_{dec} $. We can without loss of generality take the evolution of the universe after the decay of the tachyacoustic field to be radiation-dominated, since we show below that the strongest bound comes from \textit{late} decay, just at the time of Big Bang Nucleosynthesis. Assuming an instantaneous decay and saturating the limit of sound speed from Eq. (\ref{c_S_i_limit}), $ H_{l} $ is related to $ H_{dec} $ via the following two relations:
\begin{equation}
\label{H_BBN_Equation}
\ln \frac{M_P}{H_{l}}-2\ln \frac{10^{7}(1+\epsilon)^2}{\epsilon}=\left(\frac{2}{\epsilon}-1\right)\ln \frac{M_P}{H_{dec}}, 
\end{equation}
if $ H_l < H_{dec} $. Conversely, if $ H_l > H_{dec} $,
\begin{equation}
\label{H_l_large_Equation}
\ln \frac{M_P}{H_{l}}=\frac{\epsilon}{\epsilon-1}\ln \frac{10^{7}(1+\epsilon)^2}{\epsilon}.
\end{equation}
Notice that Eqs. (\ref{H_BBN_Equation}) and (\ref{H_l_large_Equation}) are the same when $ \epsilon=2 $. Due to the rapidly decreasing of speed of sound, the acoustic horizon can become shorter than the Planck length within a few e-folds. The duration before the acoustic horizon is shorter than the Planck length is given by  
\begin{equation}
\label{duration_before_sub_P_L}
\Delta N (\epsilon) =\frac{1}{\epsilon} \ln \frac{10^{7}(1+\epsilon)^2}{\epsilon}.
\end{equation}
The largest $ \Delta N $ happens when $ \epsilon \rightarrow 1 $, $ \Delta N (\epsilon \rightarrow 1) \sim 17.5 $, and $ \Delta N $ can be smaller than unity quickly as we increase $ \epsilon $, for instance $ \Delta N (\epsilon=100) \sim 0.2 $. This induces the issue of when the tachyacoustic field should decay, see Fig. \ref{fig:TC_GTCC}. Here, we consider the scenario that the tachyacoustic field is still the dominating content of the Universe even during the time when the acoustic horizon is smaller than the Planck length. However, during this period, not only the quantum-to-classical transition of modes is invalid, the perturbation calculation should break down when the length scale is shorter than the Planck length. 

Then the most relaxed constraint is that the tachyacoustic field must decay into the Standard model particles before Big Bang nucleosynthesis (BBN), so $ H_{dec} $ is a free parameter with a lower bound 
\begin{equation}
\label{H_dec_>_H_BBN}
H_{dec} \gtrsim H_{BNN},
\end{equation}
where $ H_{BBN} $ is given by the following relation from thermal equilibrium 
\begin{equation}
\label{BBN_temperature_relation_Hubble}
3M_P^2 H_{BBN}^2 =\frac{\pi^{2}}{30} g_{*}(T_{BBN} \approx 4 MeV )T_{BBN}^4.  
\end{equation}    
With  $  T_{BBN} \approx 4 MeV \sim 10^{-21}M_P $ and $ g_{*}(T_{BBN} \approx 4 MeV) \approx 10.75 $, we have $ H_{BBN} \sim 10^{-42}M_P$.
Substituting the bound from BBN, $ H_{dec} \gtrsim 10^{-42}M_P$, in 
Eq. (\ref{H_BBN_Equation}), a local minimum of $ H_{l} $ happens at the non-inflationary scenario limit, $ \epsilon \rightarrow 1 $ 
\begin{equation}
\label{H_l_BBN}
H_{l}(\epsilon \rightarrow 1) \gtrsim 10^{-57}M_P, 
\end{equation}
which is still greater than $ H_0 \approx 10^{-62}M_P $ by five orders of magnitude. Therefore, the horizon problem is unsolved under the limit $ \epsilon \rightarrow 1 $ since Eq. (\ref{H_l<H_0}) is violated.

Meanwhile, it is not difficult to see that $ H_l $ can be arbitrarily small if we allow $ \epsilon $ to be arbitrarily large in Eq. (\ref{H_BBN_Equation}), which leads to 
\begin{equation}
\label{H_l_BBN_super_epsilon}
H_{l}(\epsilon \sim 10^{45}) \gtrsim 10^{-62}M_P \sim H_0.  
\end{equation}
This was discussed in Ref. \cite{Magueijo:2008pm} that the super-Planckian energy density can be avoided  by considering an extremely large equation of state. However, this argument causes another super-Planckian issue since $ \epsilon $ and the pressure of the tachyacoustic fluid $ p $ are related by   
\begin{equation}
\label{epsilon_pressure}
\epsilon=\frac{3}{2}\left(\frac{p}{\rho}+1\right),
\end{equation}
where $ \rho $ is the energy density. 
Given that we have saturated the bound $ \rho(a_i) \sim M_P^2 H^2(a_i) \lesssim M_P^4 $ at the beginning of the tachyacoustic phase, $ \epsilon \sim 10^{45} $ means that  
\begin{equation}
\label{unreal_pressure}
p(a_i) \sim \epsilon \rho (a_i) \sim 10^{45}M_P^4,
\end{equation}
\textit{i.e.} strongly super-Planckian pressure. Therefore, for models with large $ \epsilon $, we should further consider the condition of sub-Planckian fluid pressure
\begin{equation}
\label{sub_Planck_pressure}
p(a_i) \lesssim M_P^4, 
\end{equation}   
which leads to stronger constraints on the Hubble parameter
\begin{equation}
\label{sub_Planck_pressure_Hubble}
H(a_i) \lesssim \frac{M_P}{\sqrt{\epsilon}},
\end{equation}
and the speed of sound
\begin{equation}
\label{sub_Planck_pressure_c_S}
c_S(a_i) \lesssim \frac{10^7(1+\epsilon)^2}{\epsilon^2}, 
\end{equation}     
which is approximately $10^7$ for large $ \epsilon $ and has a maximum when $ \epsilon \rightarrow 1 $.
Saturating bounds (\ref{sub_Planck_pressure_Hubble}) and (\ref{sub_Planck_pressure_c_S}), the relations between $ H_l $ and $ H_{dec} $ should be  
\begin{equation}
\label{Pressure_H_BBN_Equation}
\ln \frac{M_P}{H_{l}}-2\ln \frac{10^{7}(1+\epsilon)^2}{\epsilon^2}-\left(1-\frac{1}{\epsilon}\right)\ln \epsilon=\left(\frac{2}{\epsilon}-1\right)\ln \frac{M_P}{H_{dec}}, 
\end{equation}
if $ H_l < H_{dec} $, and 
\begin{equation}
\label{Pressure_H_l_large_Equation}
\ln \frac{M_P}{H_{l}}=\frac{\epsilon}{\epsilon-1}\ln \frac{10^{7}(1+\epsilon)^2}{\epsilon^2} + \frac{1}{2} \ln \epsilon
\end{equation}
if $ H_l > H_{dec} $,  instead of Eqs. (\ref{H_BBN_Equation}) and (\ref{H_l_large_Equation}). (Notice that Eqs. (\ref{Pressure_H_BBN_Equation}) and (\ref{Pressure_H_l_large_Equation}) reduce to Eqs. (\ref{H_BBN_Equation}) and (\ref{H_l_large_Equation}) respectively when $ \epsilon \rightarrow 1$.)   From the bound (\ref{sub_Planck_pressure_Hubble}) and the condition $ H(a_i)> H_{BBN} \sim 10^{-42}M_P $, we can obtain a bound on
$ \epsilon $ as\footnote{This actually is a relaxed bound since $ \epsilon \sim 10^{82} $ corresponds to $ H(a_i) \sim H_{BBN} $. Given that the tachyacoustic phase must end before BBN, then the duration of the tachyacoustic phase would not be enough to explain at least three decades of scale invariant curvature perturbations deduced from the CMB observation.}
\begin{equation}
\label{super_epsilon}
\epsilon <10^{82}.
\end{equation}
Even if we substitute this unrealistic bound on $ \epsilon $ with $ H_{dec} \sim H_{BBN}  $ into Eq. (\ref{Pressure_H_BBN_Equation}), we can only obtain $ H_l \sim 10^{-54}M_P $, which is greater than $ H_0 $ by eight orders of magnitude.  
Therefore, we argue that solving the horizon problem in the Tachyacoustic Cosmology unavoidably has the super-Planckian issue and the closest scenario is given by Eq. (\ref{H_l_BBN}), when $ \epsilon \rightarrow 1 $, which is short by around $ 7.5 $ e-folds, even if we consider a matter-dominated universe from $ H_l $ to $ H_0 $. 

\section{Thermal boundary conditions}

The analysis above assumes a vacuum boundary condition for perturbation modes during the tachyacoustic phase, but this is not the only possibility. In \cite{Magueijo:2008pm}, Magueijo suggests that curvature perturbations could be generated by the thermal fluctuations instead of the quantum vacuum.  In Ref. \cite{Agarwal:2014ona}, Agarwal and Afshordi consider this thermal origin of fluctuations and propose thermal Tachyacoustic Cosmology. Here we follow closely the derivations in \cite{Agarwal:2014ona}; however, we show that there is a $ 1/c_{S,i} $  factor\footnote{Notice that in \cite{Agarwal:2014ona}, the subscript $ * $ is used to represent the beginning of the tachyacoustic phase. Here, we convert it to $ i $ to match with the convention used here. We also label $ c_S(a_i) $ as $ c_{S,i} $ for simplicity in this section.} missing in Eq. (25) in \cite{Agarwal:2014ona}. By taking this factor into account, we show that thermal tachyacoustic cosmology also suffers from a super-Planckian density issue.  

In thermal Tachyacoustic Cosmology, the power spectrum of the curvature perturbation in a thermal state of temperature $ T_{i} $ is given by
\begin{equation}
\label{thermal_Power_Spec}
\langle \mathcal{P}_{\zeta}(k)\rangle_{T_{i}}= \lim_{ky\to 0^{-}}  \frac{k^3}{2\pi^2}\frac{|v_k|^2}{M_P^2 q^2}[2\langle n_{k} \rangle_{T_{i}}+1],
\end{equation}
where the thermal occupation number $ \langle n_{k} \rangle_{T_{i}} $ follows the Bose-Einstein distribution
\begin{equation}
\label{Bose_Einstein}
\langle n_{k} \rangle_{T_{i}}=\frac{1}{\exp\left(\frac{kc_{S,i}}{a_{i}T_{i}}\right)-1},
\end{equation}
and $ |v_k|^2 $ is given by the asymptotic form of Hankel function for small arguments
\begin{equation}
\label{asymptotic_Hankel}
|v_k|^2 = \frac{4^{\nu-1}\Gamma(\nu)^2}{\pi y^{2\nu-1}k^{2\nu}}+\mathcal{O}(y^{2-2\nu}).
\end{equation}
Compared to Eq. (\ref{Scale_invariant_Power_Spec_Tachy}), Eq. (\ref{thermal_Power_Spec}) has an extra term $ 2\langle n_{k} \rangle_{T_{i}}$ corresponding to the thermal fluctuations of the tachyacoustic field and $ \langle n_{k} \rangle_{T_{i}} \gg 1 $ for the thermal Tachyacoustic Cosmology. Since a thermal boundary condition for perturbations requires equilibrium, we take the background cosmology to be radiation-dominated ($\epsilon = 2$). By using Eqs. (\ref{Tachy_Forms}, \ref{q_factor}, \ref{nu}), Eq. (\ref{thermal_Power_Spec}) can be rewritten as \footnote{The parameter $\beta$ defined in \cite{Agarwal:2014ona} is related to the first flow parameter of speed of sound $s$ used here by $ \beta = -s $. } 
\begin{equation}
\label{thermal_PS_complete}
\begin{split}
\langle \mathcal{P}_{\zeta}(k)\rangle_{T_{i}}&=\frac{[-2(s+1)]^{1-\frac{3}{1+s}}}{16\pi^{3}}\Gamma \left[ 1-\frac{3}{2(s+1)}\right]^{2}\left(\frac{H_{i}}{M_{P}}\right)^2 \\ &\times \left[\frac{2}{\exp\left(\frac{kc_{S,i}}{a_{i}T_{i}}\right)-1}+1\right]
\left(\frac{k}{a_{i}H_{i}}\right)\left(\frac{kc_{S,i}}{a_{i}H_{i}}\right)^{\frac{3}{s+1}}.
\end{split}
\end{equation}
In the Rayleigh-Jeans limit $ kc_{S,i} \ll a_{i}T_{i} $, Eq. (\ref{thermal_PS_complete}) reduces to 
\begin{equation}
\label{thermal_PS_RJ_limit}
\begin{split}
\langle \mathcal{P}_{\zeta}(k)\rangle_{T_{i}}&=\frac{[-2(s+1)]^{1-\frac{3}{s+1}}}{16\pi^{3}}\Gamma \left[ 1-\frac{3}{2(s+1)}\right]^{2}\left(\frac{H_{i}}{M_{P}}\right)^2  \\
&\times \left(2\frac{T_{i}}{c_{S,i}H_{i}}\right)
\left(\frac{kc_{S,i}}{a_{i}H_{i}}\right)^{\frac{3}{s+1}},
\end{split}
\end{equation}
from which we can see that the scale invariant limit happens when $ s \rightarrow -\infty $. Then the parameter $ s $ is constrained by the measurement of the scalar spectral index $ n_s=0.965 \pm 0.004 $ \cite{Aghanim:2018eyx} as 
\begin{equation}
\label{constraint_on_s}
s=-87 \pm 10.
\end{equation} 
Under the approximation $ s \ll 0 $, we have 
\begin{equation}
\label{thermal_PS_RJ_approx}
\langle \mathcal{P}_{\zeta}(k)\rangle_{T_{i}} \approx\frac{-s}{4\pi^3}\frac{1}{M_{P}^2}\frac{H_{i}T_{i}}{c_{S,i}}\left(\frac{kc_{S,i}}{a_{i}H_{i}}\right)^{3/s}. 
\end{equation}
Following the procedure in \cite{Agarwal:2014ona}, we use Eq. (\ref{Tachy_Forms}) with $ \epsilon=2 $ and the condition of acoustic horizon crossing 
\begin{equation}
\label{acoustic_crossing}
k=\frac{a_{exit}H_{exit}}{c_S({a_{exit}})},
\end{equation}
to rewrite the last factor $ \left(\frac{kc_{S,i}}{a_{i}H_{i}}\right)^{3/s} $ in Eq. (\ref{thermal_PS_RJ_approx})
as
\begin{equation}
\label{last_factor}
\left(\frac{kc_{S,i}}{a_{i}H_{i}}\right)^{3/s} = \left(\frac{a_{i}}{a_{exit}}\right)^{3/s} \left(\frac{a_{i}}{a_{exit}}\right)^3 \approx \left(\frac{a_{i}}{a_{exit}}\right)^3, 
\end{equation}
where the condition $ s \ll 0 $ is used in the last approximation. The subscript ``exit'' represents the moment when a specific mode with comoving wavenumber $ k $ exits the acoustic horizon. Substituting Eq. (\ref{last_factor}) into Eq. (\ref{thermal_PS_RJ_approx}), we then have
\begin{equation}
\label{thermal_final_approx}
\begin{split}
\langle \mathcal{P}_{\zeta}(k)\rangle_{T_{i}} &\approx\frac{-s}{4\pi^3}\frac{H_{i}T_{i}}{M_{P}^2}\frac{1}{c_{S,i}}\left(\frac{a_{i}}{a_{exit}(k)}\right)^3\\&=\frac{-s}{4\pi^3}\frac{H_{exit}(k)T_{exit}(k)}{M_{P}^2}\frac{1}{c_{S,i}},  
\end{split}
\end{equation}
which has an extra factor $ 1/c_{S,i} $ compared to Eq. (25) in Ref. \cite{Agarwal:2014ona}. Like the case of tachyacoustic cosmology with only vacuum fluctuations, in the thermal-fluctuation-dominated scenario, the superluminal speed of sound attenuates the amplitude of the power spectrum for a given energy scale.

Now substituting the thermal equilibrium relation
\begin{equation}
\label{thermal_eqil}
H_{exit}=\frac{\pi}{\sqrt{90}}g_{*}^{1/2}\frac{T_{exit}^2}{M_P},
\end{equation} 
$ \mathcal{P}_{\zeta}(k_{*}=0.05 Mpc^{-1}) \sim 2 \times 10^{-9}$, $ g_{*} \sim 100 $, and $ s \sim -87 $ from (\ref{constraint_on_s}) into Eq. (\ref{thermal_final_approx}), we then have the relation between $ H_{exit}(k_{*}=0.05 Mpc^{-1}) $ and $ c_{S,i} $ as
\begin{equation}
\label{proof_thermal}
c_{S,i}10^{-9} \approx \left(\frac{H_{exit}(k_{*}=0.05 Mpc^{-1})}{M_P}\right)^{3/2}. 
\end{equation} 
From this relation we immediately can see that once $ c_{S,i} > 10^{9} $, the Hubble parameter is greater than the Planck mass at the time when the pivot mode $ k_{*}$ crossed the acoustic horizon (which also means that the energy scale is even larger before that moment).

To show that solving the horizon problem in this model also violates the sub-Planckian energy condition, we can substitute the approximation 
\begin{equation}
\label{Thermal_H_approximation}
H_{exit}(k_{*}) \approx H_{exit}(k_{0})
\end{equation} 
into Eq. (\ref{proof_thermal}) and saturate the condition of solving the horizon problem Eq. (\ref{general_horizon_problem}) to have
\begin{equation}
\label{thermal_SpoundSpeed_Horizon}
\frac{a_{i}H_{i}}{a_{0}H_{0}} \approx 10^{9}\left(\frac{H_i}{M_p}\right)^{3/2}.
\end{equation}
With the approximation that the universe is always radiation-dominated and $ H_{0} \sim 10^{-62}M_P$, we then have
\begin{equation}
\label{thermal_initial_H}
H_{i} \sim 10^{22}M_P,
\end{equation}
which shows the sub-Planckian energy condition is strongly violated.

\section{Conclusions}
\label{sec:Conclusions}

In this paper, we review  tachyacoustic cosmology and find that in this particular early universe model the maximum speed of sound is  limited by the amplitude of scalar power spectrum. In the scale-invariant limit, the upper bound of $ c_S $ due to the condition of sub-Planckian energy density can be written as 
\begin{equation}
c_S(a_{i}) \lesssim \frac{10^{7}(1+\epsilon)^2}{\epsilon}, 
\end{equation}
where the equality is the limit when $ H(a_i) \sim M_P $.  We also argue that considering an extremely large $ \epsilon $ to increase the bound on $ c_S $ is not viable since it corresponds to super-Planckian pressure. If we consider the condition of sub-Planckian fluid pressure, we then have stronger bounds on the Hubble parameter 
\begin{equation}
H(a_i) \lesssim \frac{M_P}{\sqrt{\epsilon}},
\end{equation}
and the speed of sound
\begin{equation}
c_S(a_i) \lesssim \frac{10^7(1+\epsilon)^2}{\epsilon^2}. 
\end{equation} 
Using these constraints we have shown that no such models can solve the horizon problem without violating either trans-Planckian energy density or pressure, and therefore, we conclude that the tachyacoustic models are not self-consistent. 

Our analysis is limited in the sense that we rule out only the case where the parameters $\epsilon$ and $s$ are constant, leading to power-law behavior for the scale factor and sound speed. In particular, we do not claim to be entirely ruling out case of superluminal sound speed in the fully general analysis of Refs. \cite{Geshnizjani:2011dk,Geshnizjani:2011rm,Geshnizjani:2013lza,Geshnizjani:2014bya}, which assumes scale invariance, but does not (as we do here) consider the normalization of the scalar power spectrum. In particular, the condition (\ref{general_horizon_problem}) depends on the assumption of approximately power-law behavior, since in the general case we cannot assume that the freezeout length of perturbations is identical to the sound horizon.  

We further examine the thermal tachyacoustic cosmology proposed in Ref. \cite{Agarwal:2014ona}, in which models involving thermal boundary conditions instead of vacuum ones. We show that in this model, the amplitude of the scalar power spectrum is also attenuated by the superluminal speed of sound, and as a result, solving the horizon problem unavoidably requires a period of super-Planckian energy density.

Relevant to recent work on ``swampland'' conjectures,  in Ref. \cite{Lin:2019pmj} we generalized the Trans-Planckian Censorship Conjecture (TCC) \cite{Bedroya:2019snp, Bedroya:2019tba} to incorporate models with a non-canonical kinetic term by proposing the Generalized Trans-Planckian Censorship Conjecture (GTCC)   
\begin{equation}
\label{GTCC}
N_{tot}<\ln\frac{c_S (a_e) M_P}{H_e},
\end{equation}  
where $ N_{tot} $ is the number of e-folds during inflation, while $ c_S (a_e) $ and $ H_e $ are the values of the speed of sound and Hubble parameter at the end of inflation respectively. This condition can trivially fit in the context of tachyacoustic cosmology, with $ N_{tot} $ as the total number of e-folds of the tachyacoustic phase. To satisfy the GTCC (\ref{GTCC}), the tachyacoustic field has to decay before the modes originated from length scale shorter than Planck length can cross the acoustic horizon (before the $ a_{GTCC} $ in Fig. \ref{fig:TC_GTCC}). This condition can further constrain the duration of the tachyacoustic phase; however, the tachyacoustic cosmology models are ruled out even without consideration of the GTCC bound. In Ref. \cite{Lin:2018edm}, we applied the distance and de Sitter swampland conjectures \cite{Obied:2018sgi, Agrawal:2018own} on tachyacoustic cosmology and found that some models can satisfy both conditions. However, in this paper we show that all of the models have trans-Planckian issues and therefore should be in the Swampland. There still remains the larger question of whether or not \textit{any} model with $c_S > 1$ should properly lie in the Swampland, a question which we leave for future work.

\section*{Acknowledgments}
This work is supported by the National Science Foundation under grant NSF-PHY-1719690. We thank Niayesh Afshordi for discussions.

\bibliographystyle{apsrev4-1}
\bibliography{TachyaCos_TP}

\begin{thebibliography}{29}%
\makeatletter
\providecommand \@ifxundefined [1]{%
 \@ifx{#1\undefined}
}%
\providecommand \@ifnum [1]{%
 \ifnum #1\expandafter \@firstoftwo
 \else \expandafter \@secondoftwo
 \fi
}%
\providecommand \@ifx [1]{%
 \ifx #1\expandafter \@firstoftwo
 \else \expandafter \@secondoftwo
 \fi
}%
\providecommand \natexlab [1]{#1}%
\providecommand \enquote  [1]{``#1''}%
\providecommand \bibnamefont  [1]{#1}%
\providecommand \bibfnamefont [1]{#1}%
\providecommand \citenamefont [1]{#1}%
\providecommand \href@noop [0]{\@secondoftwo}%
\providecommand \href [0]{\begingroup \@sanitize@url \@href}%
\providecommand \@href[1]{\@@startlink{#1}\@@href}%
\providecommand \@@href[1]{\endgroup#1\@@endlink}%
\providecommand \@sanitize@url [0]{\catcode `\\12\catcode `\$12\catcode
  `\&12\catcode `\#12\catcode `\^12\catcode `\_12\catcode `\%12\relax}%
\providecommand \@@startlink[1]{}%
\providecommand \@@endlink[0]{}%
\providecommand \url  [0]{\begingroup\@sanitize@url \@url }%
\providecommand \@url [1]{\endgroup\@href {#1}{\urlprefix }}%
\providecommand \urlprefix  [0]{URL }%
\providecommand \Eprint [0]{\href }%
\providecommand \doibase [0]{http://dx.doi.org/}%
\providecommand \selectlanguage [0]{\@gobble}%
\providecommand \bibinfo  [0]{\@secondoftwo}%
\providecommand \bibfield  [0]{\@secondoftwo}%
\providecommand \translation [1]{[#1]}%
\providecommand \BibitemOpen [0]{}%
\providecommand \bibitemStop [0]{}%
\providecommand \bibitemNoStop [0]{.\EOS\space}%
\providecommand \EOS [0]{\spacefactor3000\relax}%
\providecommand \BibitemShut  [1]{\csname bibitem#1\endcsname}%
\let\auto@bib@innerbib\@empty
\bibitem [{\citenamefont {Starobinsky}(1980)}]{Starobinsky:1980te}%
  \BibitemOpen
  \bibfield  {author} {\bibinfo {author} {\bibfnamefont {A.~A.}\ \bibnamefont
  {Starobinsky}},\ }\href {\doibase 10.1016/0370-2693(80)90670-X} {\bibfield
  {journal} {\bibinfo  {journal} {Phys. Lett.}\ }\textbf {\bibinfo {volume}
  {B91}},\ \bibinfo {pages} {99} (\bibinfo {year} {1980})},\ \bibinfo {note}
  {[,771(1980)]}\BibitemShut {NoStop}%
\bibitem [{\citenamefont {Sato}(1981{\natexlab{a}})}]{Sato:1981ds}%
  \BibitemOpen
  \bibfield  {author} {\bibinfo {author} {\bibfnamefont {K.}~\bibnamefont
  {Sato}},\ }\href {\doibase 10.1016/0370-2693(81)90805-4} {\bibfield
  {journal} {\bibinfo  {journal} {Phys. Lett.}\ }\textbf {\bibinfo {volume}
  {99B}},\ \bibinfo {pages} {66} (\bibinfo {year}
  {1981}{\natexlab{a}})}\BibitemShut {NoStop}%
\bibitem [{\citenamefont {Sato}(1981{\natexlab{b}})}]{Sato:1980yn}%
  \BibitemOpen
  \bibfield  {author} {\bibinfo {author} {\bibfnamefont {K.}~\bibnamefont
  {Sato}},\ }\href@noop {} {\bibfield  {journal} {\bibinfo  {journal} {Mon.
  Not. Roy. Astron. Soc.}\ }\textbf {\bibinfo {volume} {195}},\ \bibinfo
  {pages} {467} (\bibinfo {year} {1981}{\natexlab{b}})}\BibitemShut {NoStop}%
\bibitem [{\citenamefont {Kazanas}(1980)}]{Kazanas:1980tx}%
  \BibitemOpen
  \bibfield  {author} {\bibinfo {author} {\bibfnamefont {D.}~\bibnamefont
  {Kazanas}},\ }\href {\doibase 10.1086/183361} {\bibfield  {journal} {\bibinfo
   {journal} {Astrophys. J.}\ }\textbf {\bibinfo {volume} {241}},\ \bibinfo
  {pages} {L59} (\bibinfo {year} {1980})}\BibitemShut {NoStop}%
\bibitem [{\citenamefont {Guth}(1981)}]{Guth:1980zm}%
  \BibitemOpen
  \bibfield  {author} {\bibinfo {author} {\bibfnamefont {A.~H.}\ \bibnamefont
  {Guth}},\ }\href {\doibase 10.1103/PhysRevD.23.347} {\bibfield  {journal}
  {\bibinfo  {journal} {Phys. Rev.}\ }\textbf {\bibinfo {volume} {D23}},\
  \bibinfo {pages} {347} (\bibinfo {year} {1981})}\BibitemShut {NoStop}%
\bibitem [{\citenamefont {Linde}(1982)}]{Linde:1981mu}%
  \BibitemOpen
  \bibfield  {author} {\bibinfo {author} {\bibfnamefont {A.~D.}\ \bibnamefont
  {Linde}},\ }\bibfield  {booktitle} {\emph {\bibinfo {booktitle} {{Second
  Seminar on Quantum Gravity Moscow, USSR, October 13-15, 1981}}},\ }\href
  {\doibase 10.1016/0370-2693(82)91219-9} {\bibfield  {journal} {\bibinfo
  {journal} {Phys. Lett.}\ }\textbf {\bibinfo {volume} {108B}},\ \bibinfo
  {pages} {389} (\bibinfo {year} {1982})}\BibitemShut {NoStop}%
\bibitem [{\citenamefont {Albrecht}\ and\ \citenamefont
  {Steinhardt}(1982)}]{Albrecht:1982wi}%
  \BibitemOpen
  \bibfield  {author} {\bibinfo {author} {\bibfnamefont {A.}~\bibnamefont
  {Albrecht}}\ and\ \bibinfo {author} {\bibfnamefont {P.~J.}\ \bibnamefont
  {Steinhardt}},\ }\href {\doibase 10.1103/PhysRevLett.48.1220} {\bibfield
  {journal} {\bibinfo  {journal} {Phys. Rev. Lett.}\ }\textbf {\bibinfo
  {volume} {48}},\ \bibinfo {pages} {1220} (\bibinfo {year}
  {1982})}\BibitemShut {NoStop}%
\bibitem [{\citenamefont {Geshnizjani}\ \emph {et~al.}(2011)\citenamefont
  {Geshnizjani}, \citenamefont {Kinney},\ and\ \citenamefont
  {Moradinezhad~Dizgah}}]{Geshnizjani:2011dk}%
  \BibitemOpen
  \bibfield  {author} {\bibinfo {author} {\bibfnamefont {G.}~\bibnamefont
  {Geshnizjani}}, \bibinfo {author} {\bibfnamefont {W.~H.}\ \bibnamefont
  {Kinney}}, \ and\ \bibinfo {author} {\bibfnamefont {A.}~\bibnamefont
  {Moradinezhad~Dizgah}},\ }\href {\doibase 10.1088/1475-7516/2011/11/049}
  {\bibfield  {journal} {\bibinfo  {journal} {JCAP}\ }\textbf {\bibinfo
  {volume} {1111}},\ \bibinfo {pages} {049} (\bibinfo {year} {2011})},\ \Eprint
  {http://arxiv.org/abs/1107.1241} {arXiv:1107.1241 [astro-ph.CO]} \BibitemShut
  {NoStop}%
\bibitem [{\citenamefont {Geshnizjani}\ \emph {et~al.}(2012)\citenamefont
  {Geshnizjani}, \citenamefont {Kinney},\ and\ \citenamefont
  {Moradinezhad~Dizgah}}]{Geshnizjani:2011rm}%
  \BibitemOpen
  \bibfield  {author} {\bibinfo {author} {\bibfnamefont {G.}~\bibnamefont
  {Geshnizjani}}, \bibinfo {author} {\bibfnamefont {W.~H.}\ \bibnamefont
  {Kinney}}, \ and\ \bibinfo {author} {\bibfnamefont {A.}~\bibnamefont
  {Moradinezhad~Dizgah}},\ }\href {\doibase 10.1088/1475-7516/2012/02/015}
  {\bibfield  {journal} {\bibinfo  {journal} {JCAP}\ }\textbf {\bibinfo
  {volume} {1202}},\ \bibinfo {pages} {015} (\bibinfo {year} {2012})},\ \Eprint
  {http://arxiv.org/abs/1110.4640} {arXiv:1110.4640 [astro-ph.CO]} \BibitemShut
  {NoStop}%
\bibitem [{\citenamefont {Geshnizjani}\ and\ \citenamefont
  {Ahmadi}(2013)}]{Geshnizjani:2013lza}%
  \BibitemOpen
  \bibfield  {author} {\bibinfo {author} {\bibfnamefont {G.}~\bibnamefont
  {Geshnizjani}}\ and\ \bibinfo {author} {\bibfnamefont {N.}~\bibnamefont
  {Ahmadi}},\ }\href {\doibase 10.1088/1475-7516/2013/11/029} {\bibfield
  {journal} {\bibinfo  {journal} {JCAP}\ }\textbf {\bibinfo {volume} {1311}},\
  \bibinfo {pages} {029} (\bibinfo {year} {2013})},\ \Eprint
  {http://arxiv.org/abs/1309.4782} {arXiv:1309.4782 [hep-th]} \BibitemShut
  {NoStop}%
\bibitem [{\citenamefont {Geshnizjani}\ and\ \citenamefont
  {Kinney}(2015)}]{Geshnizjani:2014bya}%
  \BibitemOpen
  \bibfield  {author} {\bibinfo {author} {\bibfnamefont {G.}~\bibnamefont
  {Geshnizjani}}\ and\ \bibinfo {author} {\bibfnamefont {W.~H.}\ \bibnamefont
  {Kinney}},\ }\href {\doibase 10.1088/1475-7516/2015/08/008} {\bibfield
  {journal} {\bibinfo  {journal} {JCAP}\ }\textbf {\bibinfo {volume} {1508}},\
  \bibinfo {pages} {008} (\bibinfo {year} {2015})},\ \Eprint
  {http://arxiv.org/abs/1410.4968} {arXiv:1410.4968 [astro-ph.CO]} \BibitemShut
  {NoStop}%
\bibitem [{\citenamefont {Bessada}\ \emph {et~al.}(2010)\citenamefont
  {Bessada}, \citenamefont {Kinney}, \citenamefont {Stojkovic},\ and\
  \citenamefont {Wang}}]{Bessada:2009ns}%
  \BibitemOpen
  \bibfield  {author} {\bibinfo {author} {\bibfnamefont {D.}~\bibnamefont
  {Bessada}}, \bibinfo {author} {\bibfnamefont {W.~H.}\ \bibnamefont {Kinney}},
  \bibinfo {author} {\bibfnamefont {D.}~\bibnamefont {Stojkovic}}, \ and\
  \bibinfo {author} {\bibfnamefont {J.}~\bibnamefont {Wang}},\ }\href {\doibase
  10.1103/PhysRevD.81.043510} {\bibfield  {journal} {\bibinfo  {journal} {Phys.
  Rev.}\ }\textbf {\bibinfo {volume} {D81}},\ \bibinfo {pages} {043510}
  (\bibinfo {year} {2010})},\ \Eprint {http://arxiv.org/abs/0908.3898}
  {arXiv:0908.3898 [astro-ph.CO]} \BibitemShut {NoStop}%
\bibitem [{\citenamefont {Magueijo}(2008)}]{Magueijo:2008pm}%
  \BibitemOpen
  \bibfield  {author} {\bibinfo {author} {\bibfnamefont {J.}~\bibnamefont
  {Magueijo}},\ }\href {\doibase 10.1103/PhysRevLett.100.231302} {\bibfield
  {journal} {\bibinfo  {journal} {Phys. Rev. Lett.}\ }\textbf {\bibinfo
  {volume} {100}},\ \bibinfo {pages} {231302} (\bibinfo {year} {2008})},\
  \Eprint {http://arxiv.org/abs/0803.0859} {arXiv:0803.0859 [astro-ph]}
  \BibitemShut {NoStop}%
\bibitem [{\citenamefont {Babichev}\ \emph {et~al.}(2008)\citenamefont
  {Babichev}, \citenamefont {Mukhanov},\ and\ \citenamefont
  {Vikman}}]{Babichev:2007dw}%
  \BibitemOpen
  \bibfield  {author} {\bibinfo {author} {\bibfnamefont {E.}~\bibnamefont
  {Babichev}}, \bibinfo {author} {\bibfnamefont {V.}~\bibnamefont {Mukhanov}},
  \ and\ \bibinfo {author} {\bibfnamefont {A.}~\bibnamefont {Vikman}},\ }\href
  {\doibase 10.1088/1126-6708/2008/02/101} {\bibfield  {journal} {\bibinfo
  {journal} {JHEP}\ }\textbf {\bibinfo {volume} {02}},\ \bibinfo {pages} {101}
  (\bibinfo {year} {2008})},\ \Eprint {http://arxiv.org/abs/0708.0561}
  {arXiv:0708.0561 [hep-th]} \BibitemShut {NoStop}%
\bibitem [{\citenamefont {Kinney}(2002)}]{Kinney:2002qn}%
  \BibitemOpen
  \bibfield  {author} {\bibinfo {author} {\bibfnamefont {W.~H.}\ \bibnamefont
  {Kinney}},\ }\href {\doibase 10.1103/PhysRevD.66.083508} {\bibfield
  {journal} {\bibinfo  {journal} {Phys. Rev.}\ }\textbf {\bibinfo {volume}
  {D66}},\ \bibinfo {pages} {083508} (\bibinfo {year} {2002})},\ \Eprint
  {http://arxiv.org/abs/astro-ph/0206032} {arXiv:astro-ph/0206032 [astro-ph]}
  \BibitemShut {NoStop}%
\bibitem [{\citenamefont {Bean}\ \emph {et~al.}(2008)\citenamefont {Bean},
  \citenamefont {Chung},\ and\ \citenamefont {Geshnizjani}}]{Bean:2008ga}%
  \BibitemOpen
  \bibfield  {author} {\bibinfo {author} {\bibfnamefont {R.}~\bibnamefont
  {Bean}}, \bibinfo {author} {\bibfnamefont {D.~J.~H.}\ \bibnamefont {Chung}},
  \ and\ \bibinfo {author} {\bibfnamefont {G.}~\bibnamefont {Geshnizjani}},\
  }\href {\doibase 10.1103/PhysRevD.78.023517} {\bibfield  {journal} {\bibinfo
  {journal} {Phys. Rev.}\ }\textbf {\bibinfo {volume} {D78}},\ \bibinfo {pages}
  {023517} (\bibinfo {year} {2008})},\ \Eprint {http://arxiv.org/abs/0801.0742}
  {arXiv:0801.0742 [astro-ph]} \BibitemShut {NoStop}%
\bibitem [{\citenamefont {Lin}\ and\ \citenamefont
  {Kinney}(2019{\natexlab{a}})}]{Lin:2018edm}%
  \BibitemOpen
  \bibfield  {author} {\bibinfo {author} {\bibfnamefont {W.-C.}\ \bibnamefont
  {Lin}}\ and\ \bibinfo {author} {\bibfnamefont {W.~H.}\ \bibnamefont
  {Kinney}},\ }\href {\doibase 10.1088/1475-7516/2019/10/038} {\bibfield
  {journal} {\bibinfo  {journal} {JCAP}\ }\textbf {\bibinfo {volume} {1910}},\
  \bibinfo {pages} {038} (\bibinfo {year} {2019}{\natexlab{a}})},\ \Eprint
  {http://arxiv.org/abs/1812.04447} {arXiv:1812.04447 [astro-ph.CO]}
  \BibitemShut {NoStop}%
\bibitem [{\citenamefont {Garriga}\ and\ \citenamefont
  {Mukhanov}(1999)}]{Garriga:1999vw}%
  \BibitemOpen
  \bibfield  {author} {\bibinfo {author} {\bibfnamefont {J.}~\bibnamefont
  {Garriga}}\ and\ \bibinfo {author} {\bibfnamefont {V.~F.}\ \bibnamefont
  {Mukhanov}},\ }\href {\doibase 10.1016/S0370-2693(99)00602-4} {\bibfield
  {journal} {\bibinfo  {journal} {Phys. Lett.}\ }\textbf {\bibinfo {volume}
  {B458}},\ \bibinfo {pages} {219} (\bibinfo {year} {1999})},\ \Eprint
  {http://arxiv.org/abs/hep-th/9904176} {arXiv:hep-th/9904176 [hep-th]}
  \BibitemShut {NoStop}%
\bibitem [{\citenamefont {Peiris}\ \emph {et~al.}(2007)\citenamefont {Peiris},
  \citenamefont {Baumann}, \citenamefont {Friedman},\ and\ \citenamefont
  {Cooray}}]{Peiris:2007gz}%
  \BibitemOpen
  \bibfield  {author} {\bibinfo {author} {\bibfnamefont {H.~V.}\ \bibnamefont
  {Peiris}}, \bibinfo {author} {\bibfnamefont {D.}~\bibnamefont {Baumann}},
  \bibinfo {author} {\bibfnamefont {B.}~\bibnamefont {Friedman}}, \ and\
  \bibinfo {author} {\bibfnamefont {A.}~\bibnamefont {Cooray}},\ }\href
  {\doibase 10.1103/PhysRevD.76.103517} {\bibfield  {journal} {\bibinfo
  {journal} {Phys. Rev.}\ }\textbf {\bibinfo {volume} {D76}},\ \bibinfo {pages}
  {103517} (\bibinfo {year} {2007})},\ \Eprint {http://arxiv.org/abs/0706.1240}
  {arXiv:0706.1240 [astro-ph]} \BibitemShut {NoStop}%
\bibitem [{\citenamefont {Kinney}\ and\ \citenamefont
  {Tzirakis}(2008)}]{Kinney:2007ag}%
  \BibitemOpen
  \bibfield  {author} {\bibinfo {author} {\bibfnamefont {W.~H.}\ \bibnamefont
  {Kinney}}\ and\ \bibinfo {author} {\bibfnamefont {K.}~\bibnamefont
  {Tzirakis}},\ }\href {\doibase 10.1103/PhysRevD.77.103517} {\bibfield
  {journal} {\bibinfo  {journal} {Phys. Rev.}\ }\textbf {\bibinfo {volume}
  {D77}},\ \bibinfo {pages} {103517} (\bibinfo {year} {2008})},\ \Eprint
  {http://arxiv.org/abs/0712.2043} {arXiv:0712.2043 [astro-ph]} \BibitemShut
  {NoStop}%
\bibitem [{\citenamefont {Khoury}\ and\ \citenamefont
  {Piazza}(2009)}]{Khoury:2008wj}%
  \BibitemOpen
  \bibfield  {author} {\bibinfo {author} {\bibfnamefont {J.}~\bibnamefont
  {Khoury}}\ and\ \bibinfo {author} {\bibfnamefont {F.}~\bibnamefont
  {Piazza}},\ }\href {\doibase 10.1088/1475-7516/2009/07/026} {\bibfield
  {journal} {\bibinfo  {journal} {JCAP}\ }\textbf {\bibinfo {volume} {0907}},\
  \bibinfo {pages} {026} (\bibinfo {year} {2009})},\ \Eprint
  {http://arxiv.org/abs/0811.3633} {arXiv:0811.3633 [hep-th]} \BibitemShut
  {NoStop}%
\bibitem [{\citenamefont {Bessada}\ and\ \citenamefont
  {Kinney}(2012)}]{Bessada:2012zx}%
  \BibitemOpen
  \bibfield  {author} {\bibinfo {author} {\bibfnamefont {D.}~\bibnamefont
  {Bessada}}\ and\ \bibinfo {author} {\bibfnamefont {W.~H.}\ \bibnamefont
  {Kinney}},\ }\href {\doibase 10.1103/PhysRevD.86.083502} {\bibfield
  {journal} {\bibinfo  {journal} {Phys. Rev.}\ }\textbf {\bibinfo {volume}
  {D86}},\ \bibinfo {pages} {083502} (\bibinfo {year} {2012})},\ \Eprint
  {http://arxiv.org/abs/1206.2711} {arXiv:1206.2711 [gr-qc]} \BibitemShut
  {NoStop}%
\bibitem [{\citenamefont {Lin}\ and\ \citenamefont
  {Kinney}(2019{\natexlab{b}})}]{Lin:2019pmj}%
  \BibitemOpen
  \bibfield  {author} {\bibinfo {author} {\bibfnamefont {W.-C.}\ \bibnamefont
  {Lin}}\ and\ \bibinfo {author} {\bibfnamefont {W.~H.}\ \bibnamefont
  {Kinney}},\ }\href@noop {} {\  (\bibinfo {year} {2019}{\natexlab{b}})},\
  \Eprint {http://arxiv.org/abs/1911.03736} {arXiv:1911.03736 [gr-qc]}
  \BibitemShut {NoStop}%
\bibitem [{\citenamefont {Aghanim}\ \emph {et~al.}(2018)\citenamefont {Aghanim}
  \emph {et~al.}}]{Aghanim:2018eyx}%
  \BibitemOpen
  \bibfield  {author} {\bibinfo {author} {\bibfnamefont {N.}~\bibnamefont
  {Aghanim}} \emph {et~al.} (\bibinfo {collaboration} {Planck}),\ }\href@noop
  {} {\  (\bibinfo {year} {2018})},\ \Eprint {http://arxiv.org/abs/1807.06209}
  {arXiv:1807.06209 [astro-ph.CO]} \BibitemShut {NoStop}%
\bibitem [{\citenamefont {Agarwal}\ and\ \citenamefont
  {Afshordi}(2014)}]{Agarwal:2014ona}%
  \BibitemOpen
  \bibfield  {author} {\bibinfo {author} {\bibfnamefont {A.}~\bibnamefont
  {Agarwal}}\ and\ \bibinfo {author} {\bibfnamefont {N.}~\bibnamefont
  {Afshordi}},\ }\href {\doibase 10.1103/PhysRevD.90.043528} {\bibfield
  {journal} {\bibinfo  {journal} {Phys. Rev. D}\ }\textbf {\bibinfo {volume}
  {90}},\ \bibinfo {pages} {043528} (\bibinfo {year} {2014})},\ \Eprint
  {http://arxiv.org/abs/1406.0575} {arXiv:1406.0575 [astro-ph.CO]} \BibitemShut
  {NoStop}%
\bibitem [{\citenamefont {Bedroya}\ and\ \citenamefont
  {Vafa}(2019)}]{Bedroya:2019snp}%
  \BibitemOpen
  \bibfield  {author} {\bibinfo {author} {\bibfnamefont {A.}~\bibnamefont
  {Bedroya}}\ and\ \bibinfo {author} {\bibfnamefont {C.}~\bibnamefont {Vafa}},\
  }\href@noop {} {\  (\bibinfo {year} {2019})},\ \Eprint
  {http://arxiv.org/abs/1909.11063} {arXiv:1909.11063 [hep-th]} \BibitemShut
  {NoStop}%
\bibitem [{\citenamefont {Bedroya}\ \emph {et~al.}(2019)\citenamefont
  {Bedroya}, \citenamefont {Brandenberger}, \citenamefont {Loverde},\ and\
  \citenamefont {Vafa}}]{Bedroya:2019tba}%
  \BibitemOpen
  \bibfield  {author} {\bibinfo {author} {\bibfnamefont {A.}~\bibnamefont
  {Bedroya}}, \bibinfo {author} {\bibfnamefont {R.}~\bibnamefont
  {Brandenberger}}, \bibinfo {author} {\bibfnamefont {M.}~\bibnamefont
  {Loverde}}, \ and\ \bibinfo {author} {\bibfnamefont {C.}~\bibnamefont
  {Vafa}},\ }\href@noop {} {\  (\bibinfo {year} {2019})},\ \Eprint
  {http://arxiv.org/abs/1909.11106} {arXiv:1909.11106 [hep-th]} \BibitemShut
  {NoStop}%
\bibitem [{\citenamefont {Obied}\ \emph {et~al.}(2018)\citenamefont {Obied},
  \citenamefont {Ooguri}, \citenamefont {Spodyneiko},\ and\ \citenamefont
  {Vafa}}]{Obied:2018sgi}%
  \BibitemOpen
  \bibfield  {author} {\bibinfo {author} {\bibfnamefont {G.}~\bibnamefont
  {Obied}}, \bibinfo {author} {\bibfnamefont {H.}~\bibnamefont {Ooguri}},
  \bibinfo {author} {\bibfnamefont {L.}~\bibnamefont {Spodyneiko}}, \ and\
  \bibinfo {author} {\bibfnamefont {C.}~\bibnamefont {Vafa}},\ }\href@noop {}
  {\  (\bibinfo {year} {2018})},\ \Eprint {http://arxiv.org/abs/1806.08362}
  {arXiv:1806.08362 [hep-th]} \BibitemShut {NoStop}%
\bibitem [{\citenamefont {Agrawal}\ \emph {et~al.}(2018)\citenamefont
  {Agrawal}, \citenamefont {Obied}, \citenamefont {Steinhardt},\ and\
  \citenamefont {Vafa}}]{Agrawal:2018own}%
  \BibitemOpen
  \bibfield  {author} {\bibinfo {author} {\bibfnamefont {P.}~\bibnamefont
  {Agrawal}}, \bibinfo {author} {\bibfnamefont {G.}~\bibnamefont {Obied}},
  \bibinfo {author} {\bibfnamefont {P.~J.}\ \bibnamefont {Steinhardt}}, \ and\
  \bibinfo {author} {\bibfnamefont {C.}~\bibnamefont {Vafa}},\ }\href {\doibase
  10.1016/j.physletb.2018.07.040} {\bibfield  {journal} {\bibinfo  {journal}
  {Phys. Lett.}\ }\textbf {\bibinfo {volume} {B784}},\ \bibinfo {pages} {271}
  (\bibinfo {year} {2018})},\ \Eprint {http://arxiv.org/abs/1806.09718}
  {arXiv:1806.09718 [hep-th]} \BibitemShut {NoStop}%
\end{thebibliography}%


\end{document}